\documentclass[a4paper,amsmath,amssymb,12pt]{article}

\def \be {\begin{equation}}
\def \ee {\end{equation}}
\def \bea {\begin{eqnarray}}
\def \eea {\end{eqnarray}}
\def \nn {\nonumber}

\def \a {\alpha}
\def \b {\beta}

\def \d {\delta}

\def \m {\mu}
\def \n {\nu}
\def \k {\kappa}

\def \s {\sigma}
\def \r {\rho}
\def \o {\omega}

\def \th {\theta}
\def \Th {\Theta}

\def \t {\tau}
\def \dag {\dagger}
\def \p {\partial}

\def\bd{\begin{document}}
\def\ed{\end{document}}
\def\nn{\nonumber}
\def\bea{\begin{eqnarray}}
\def\eea{\end{eqnarray}}
\let\bm=\bibitem
\let\la=\label

\def\N{{\cal N}}
\def\sst{\scriptscriptstyle}
\def\thetabar{\bar\theta}
\def\Tr{{\rm Tr}}
\def\one{\mbox{1 \kern-.59em {\rm l}}}

%

\def\a{\alpha}      \def\da{{\dot\alpha}}
\def\b{\beta}       \def\db{{\dot\beta}}
\def\c{\gamma}  \def\C{\Gamma}  \def\cdt{\dot\gamma}
\def\d{\delta}  \def\D{\Delta}  \def\ddt{\dot\delta}
\def\e{\epsilon}        \def\vare{\varepsilon}
\def\f{\phi}    \def\F{\Phi}    \def\vvf{\f}
\def\h{\eta}
\def\k{\kappa}
\def\l{\lambda} \def\L{\Lambda}
\def\m{\mu} \def\n{\nu}
\def\o{\omega}
\def\P{\Pi}
\def\r{\rho}
\def\s{\sigma}  \def\S{\Sigma}
\def\t{\tau}
\def\th{\theta} \def\Th{\Theta} \def\vth{\vartheta}
\def\X{\Xeta}
\def\z{\zeta}
\def\w{\wedge}
\def\u{\underline}
\def\hs{\hspace}


\def\cA{{\cal A}} \def\cB{{\cal B}} \def\cC{{\cal C}}
\def\cD{{\cal D}} \def\cE{{\cal E}} \def\cF{{\cal F}}
\def\cG{{\cal G}} \def\cH{{\cal H}} \def\cI{{\cal I}}
\def\cJ{{\cal J}} \def\cK{{\cal K}} \def\cL{{\cal L}}
\def\cM{{\cal M}} \def\cN{{\cal N}} \def\cO{{\cal O}}
\def\cP{{\cal P}} \def\cQ{{\cal Q}} \def\cR{{\cal R}}
\def\cS{{\cal S}} \def\cT{{\cal T}} \def\cU{{\cal U}}
\def\cV{{\cal V}} \def\cW{{\cal W}} \def\cX{{\cal X}}
\def\cY{{\cal Y}} \def\cZ{{\cal Z}}


\def\ua{\underline{\alpha}} \def\ubb{\underline{\beta}}
\def\ug{\underline{\gamma}}
\def\ub{\underline{\phantom{\alpha}}\!\!\!\beta}
\def\uc{\underline{\phantom{\alpha}}\!\!\!\gamma}
\def\um{\underline{\mu}} \def\un{\underline{\nu}}
\def\ud{\underline\delta}
\def\ue{\underline\epsilon}
\def\una{\underline a}\def\unA{\underline A}
\def\unb{\underline b}\def\unB{\underline B}
\def\unc{\underline c}\def\unC{\underline C}
\def\und{\underline d}\def\unD{\underline D}
\def\une{\underline e}\def\unE{\underline E}
\def\unf{\underline{\phantom{e}}\!\!\!\! f}\def\unF{\underline F}
\def\unm{\underline m}\def\unM{\underline M}
\def\unn{\underline n}\def\unN{\underline N}
\def\unp{\underline{\phantom{a}}\!\!\! p}
\def\unP{\underline P}
\def\unq{\underline{\phantom{a}}\!\!\! q}
\def\unQ{\underline{\phantom{A}}\!\!\!\! Q}
\def\unH{\underline{H}}
\def\ul{\underline}

\def\As {{A \hspace{-6.4pt} \slash}\;}
\def\bs {{b \hspace{-6.4pt} \slash}\;}
\def\Ds {{D \hspace{-6.4pt} \slash}\;}
\def\ds {{\del \hspace{-6.4pt} \slash}\;}
\def\ss {{\s \hspace{-6.4pt} \slash}\;}
\def\ks {{ k \hspace{-6.4pt} \slash}\;}
\def\ps {{p \hspace{-6.4pt} \slash}\;}
\def\pas {{{p_1} \hspace{-6.4pt} \slash}\;}
\def\pbs {{{p_2} \hspace{-6.4pt} \slash}\;}


\def\Fh{\hat{F}}
\def\Vh{\hat{V}}
\def\Xh{\hat{X}}
\def\ah{\hat{a}}
\def\xh{\hat{x}}
\def\yh{\hat{y}}
\def\ph{\hat{p}}
\def\xih{\hat{\xi}}

\def\psit{\tilde{\psi}}
\def\Psit{\tilde{\Psi}}
\def\tht{\tilde{\th}}

\def\At{\tilde{A}}
\def\Qt{\tilde{Q}}
\def\Rt{\tilde{R}}
\def\Nt{\tilde{N}}

\def\at{\tilde{a}}
\def\st{\tilde{s}}
\def\ft{\tilde{f}}
\def\pt{\tilde{p}}
\def\qt{\tilde{q}}
\def\vt{\tilde{v}}
\def\nt{\tilde{n}}


\def\delb{\bar{\partial}}
\def\bz{\bar{z}}
\def\bD{\bar{D}}
\def\bB{\bar{B}}


\def\bk{{\bf k}}
\def\bl{{\bf l}}
\def\bp{{\bf p}}
\def\bq{{\bf q}}
\def\br{{\bf r}}
\def\bx{{\bf x}}
\def\by{{\bf y}}
\def\bR{{\bf R}}
\def\bV{{\bf V}}


\def\d{\delta}\def\D{\Delta}\def\ddt{\dot\delta}

\def\p{\partial} \def\del{\partial}
\def\xx{\times}
\def\uno{\mbox{1 \kern-.59em {\rm l}}}

\def\trp{^{\top}}
\def\inv{^{-1}}
\def\dag{{^{\dagger}}}
\def\pr{\prime}
\textwidth=6.0in \hoffset=-.3in \textheight=9in \voffset=-.8in
\def\baselinestretch{1.2}

\def\rar{\rightarrow}
\def\lar{\leftarrow}
\def\lrar{\leftrightarrow}

\title{Quasinormal modes of warped $AdS_3$ black holes and warped AdS/CFT correspondence}
\author{Bin Chen and Zhi-bo Xu\footnote{Email:bchen01,xuzhibo@pku.edu.cn}\\
{\small Department of Physics} \\
{\small and State Key Laboratory of Nuclear Physics and Technology,}\\
{\small Peking University, Beijing 100871, P.R.China}}
\date{}
\bd \maketitle
\begin{abstract}
We calculate the scalar quasinormal modes of warped $AdS_3$ black
holes analytically. We find that in general they are not in
agreement with the prediction of the usual AdS/CFT correspondence.
Nevertheless, for the black hole not deviating far from BTZ black
hole, the quasinormal frequencies are still well consistent with
AdS/CFT correspondence.
\end{abstract}
\newpage

The quasinormal modes of black holes have been studied in general
relativity for nearly forty years (please see \cite{QNM} for
excellent reviews and detailed list of references). They
characterize the ``sound" of the black holes. For  the
perturbations of the black holes, they obey the linearized
equations of motion. And the quasinormal modes are defined as the
perturbations  subject to the physical boundary conditions that
near the horizon of the black holes the local solution is purely
ingoing and at spatial infinity the solution is purely outgoing.
This choice of boundary condition makes generically the frequency
of the perturbation be complex. Therefore the quasinormal
perturbations undergo damped oscillations, just as the ring of a
bell. Physically, the perturbations can fall into the black holes
and decay from the point of view of an observer at the infinity.
This is true not only for the black holes in flat spacetime and
also in anti-de Sitter(AdS) spacetime.

On the other hand, the quasinormal modes of AdS black holes are of
particular interests in the study of AdS/CFT
correspondence\cite{AdSCFT}. From AdS/CFT correspondence, the
presence of black holes in anti-de Sitter spacetime corresponds to
turning on a temperature in dual conformal field theory. The
quasinormal modes of black holes correspond to the tiny deviations
of the thermal equilibrium in dual field theory, which have
nonzero damping\cite{Horowitz99}. It turns out that the complex
frequencies of quasinormal modes can be understood as the poles in
the retarded green function in the dual finite temperature field
theory\cite{Birmingham01, Son05}. Especially for two-dimensional
conformal field theory, the left and right sectors are
independent. At thermal equilibrium, the two sectors
 may have different temperatures $(T_L,T_R)$. Consider a small perturbation
 operator ${\cal O}$ with conformal weights $(h_L, h_R)$. Under such a
 perturbation, the system will return to thermal equilibrium
 exponentially with a characteristic time scale, which is
 inversely proportional to the imaginary part of the poles of the
 correlation function of the operator ${\cal O}$ in momentum
 space. There are two sets of poles
 \bea\label{pole}
 \o_L&=&k-4\pi iT_L(n+h_L) \nn\\
 \o_R&=&-k-4\pi iT_R(n+h_R),
 \eea
where $n$ being non-negative integer. In \cite{Birmingham01}, it
has been shown that these poles are in exact agreement with the
quasinormal frequencies of the BTZ black hole.

The Banados-Teitelboim-Zanelli(BTZ) black hole is a solution of
the vacuum Einstein equations in three-dimensional anti-de Sitter
spacetime\cite{BTZ}. The metric of the BTZ black hole is of the
form
 \bea\label{BTZ}
 ds^2&=&-\big(-M+\frac{\rho^2}{l^2}+\frac{J^2}{4\rho^2}\big)d\tau^2+
 \big(-M+\frac{\rho^2}{l^2}+\frac{J^2}{4\rho^2}\big)^{-1}d\rho^2\nn\\
 & &+\rho^2\big(d\phi-\frac{J}{2\rho^2}d\tau\big)^2,
 \eea
 in Schwarzschild coordinates, where $l$ is the radius of $AdS_3$ spacetime and $M, J$ are the
 mass and angular momentum of the black hole. There are two
 horizons in BTZ black hole, outer one $\rho_+$ and inner one $\rho_-$
 decided by $M,J$. BTZ black holes are asymptotic to $AdS_3$ spacetime.
 According to $AdS_3/CFT_2$
 dictionary, the two independent temperatures are related to the
 horizons
 \be\label{tempBTZ}
 T_L=\frac{\rho_+-\rho_-}{2\pi l}, \hspace{5ex} T_R=\frac{\rho_++\rho_-}{2\pi
 l}.
 \ee
 For a scalar field with mass $m$ in $AdS_3$, its corresponding
 operator in $CFT_2$ has conformal weights
 \be
 h_L=h_R=\frac{1}{2}(1+\sqrt{1+m^2l^2}).
 \ee
 The quasinormal modes of the BTZ black hole were
computed for the first time in \cite{Cardoso}. And it was found in
\cite{Birmingham01} that the scalar quasinormal frequencies of BTZ
black hole were in precise
 match with (\ref{pole}).

The warped $AdS_3$ black hole is the vacuum solution in three
dimensional topologically massive gravity (TMG) with a negative
cosmological constant. It was first discovered in \cite{BC07}.
Very recently, it was found that just as BTZ black holes are
discrete quotients of $AdS_3$ spacetime, the warped $AdS_3$ black
holes are discrete quotients of warped $AdS_3$
spacetime\cite{Andy08}. It was even conjectured in \cite{Andy08}
that the warped black holes are holographically dual to a
two-dimensional conformal field theory. The study of
thermodynamics of the warped black holes gave strong support of
the conjecture. It would be interesting to investigate this
 nontrivial $AdS_3/CFT_2$ correspondence further\footnote{It
 would be better to call it as warped $AdS_3/CFT_2$ correspondence, since the
 conformal boundary of warped $AdS_3$ is different from the one of $AdS_3$.
 We thank the referee for this suggestion.}. For
other recent works on warped $AdS_3$ black holes, see
\cite{warped}.

In this letter, we study the scalar quasinormal modes of the
warped $AdS_3$ black holes and check if it is consistent with the
prediction of AdS/CFT correspondence.

Let us start from the spacelike stretched black holes, discussed
in \cite{BC07,Andy08}. Its metric takes the following form in
terms of Schwarzschild coordinates:
 \bea\label{warped}
 ds^2=-N(r)^2dt^2+R(r)^2[d\th +N^\th
 (r)dt]^2+\frac{l^2dr^2}{4R(r)^2N(r)^2},
 \eea
 where
 \bea
 R(r)^2&=&\frac{r}{4}\left(3(v^2-1)r+(v^2+3)(r_++r_-)-4v\sqrt{r_+r_-(v^2+3)}\right)
 \\
 N^2(r)&=&\frac{(v^2+3)(r-r_+)(r-r_-)}{4R(r)^2},\\
 N^\th(r)&=&\frac{2vr-\sqrt{r_+r_-(v^2+3)}}{2R(r)^2},
 \eea
 where $-l^{-2}$ is the negative cosmological constant and the parameter
 $v=\mu l/3$ with $\mu$ being the mass of the graviton.
 Just like the BTZ black hole, there are two horizons located at
 $r=r_+$ and $r=r_-$. We will focus on the physical black holes, in
 which case $v\geq 1$. When $v=1$, there is no stretching and
 the above black hole becomes the usual BTZ black hole. In
 \cite{Andy08}, the temperatures of the black holes were identified
 to be
  \be
   T_H=\frac{4\pi vl}{v^2+3}\frac{T_L+T_R}{T_R},
 \ee
 where
  \bea\label{tempwarped}
  T_L&=&\frac{(v^2+3)}{8\pi
  l}\left(r_++r_--\frac{\sqrt{(v^2+3)r_+r_-}}{v}\right) \\
  T_R&=&\frac{(v^2+3)(r_+-r_-)}{8\pi
  l}
 \eea
 are the temperature of dual CFT. The dual two-dimensional CFT is supposed to have the central
 charges
  \be
  c_L=\frac{l}{G}\frac{4v}{v^2+3}, \hspace{5ex}
  c_R=\frac{l}{G}\frac{5v^2+3}{v(v^2+3)}.
  \ee

When $v=1$, the warped black holes is the BTZ black holes. This
could be seen by a coordinate transformation
 \be\label{time}
 t=\frac{\rho_+-\rho_-}{l}\tau, \hspace{3ex}
 \th=\phi-\frac{1}{l}\tau,  \hspace{3ex}
 r=\frac{\rho^2}{\rho_+-\rho_-}.
 \ee
In terms of the coordinates $r$, the horizons are
 \be
 \rho_\pm=\sqrt{r_\pm}(\sqrt{r_+}-\sqrt{r_-}).
 \ee
Note that the definition of the temperatures is actually related
to the coordinate choice. In BTZ case, one has
 \bea
 T_L&=&\frac{\rho_+-\rho_-}{2\pi l}=\frac{(\sqrt{r_+}-\sqrt{r_-})^2}{2\pi l}
 \nn\\
 T_R&=&\frac{\rho_++\rho_-}{2\pi l}=\frac{{r_+}-{r_-}}{2\pi l},\nn
 \eea
which is the same as (\ref{tempwarped}). However, notice that the
transformation (\ref{time}) relate two time direction in two
different coordinates so that the temperatures in two coordinates
should be related by
 \be
 T(r)=\frac{l}{\rho_+-\rho_-}T(\rho).
 \ee
 Taking into account of this subtlety, one has
 \bea\label{tempr}
 \tilde{T}_L&=&\frac{1}{2\pi},\nn \\
 \tilde{T}_R&=&\frac{1}{2\pi}\frac{\sqrt{r_+}+\sqrt{r_-}}{\sqrt{r_+}-\sqrt{r_-}}.
 \eea

 To study the quasinormal modes of the blackhole, we
 consider the scalar perturbation in the background for simplicity. The scalar
 wavefunction satisfies the equation
  \be
  (\nabla_\mu\nabla^\mu-m^2) \Phi=0
  \ee
  Since the background has the translational isometry along $t$
  and $\th$, we may make the following ansatz
  \be
  \Phi=e^{-i\o t+ik\th}\phi,
  \ee
to simplify the
  equation, and we have
  \be
  \frac{4}{l^2}R^2N^2\p_r(R^2N^2\p_r)\phi+(R^2\o^2+2k(R^2N^\th)\o+k^2-m^2N^2R^2)\phi=0.
  \ee
Define the variable \be z=\frac{r-r_+}{r-r_-}, \ee and we are led
to the following equation after a lengthy calculation
 \be\label{radial}
 z(1-z)\frac{d^2\phi}{dz^2}+(1-z)\frac{d\phi}{dz}+\frac{1}{(v^2+3)^2}\left(\frac{A}{z}+B+\frac{C}{1-z}\right)\phi=0,
 \ee
where
 \bea
 A&=&\frac{l^2}{(r_+-r_-)^2}\big(2k+\o\sqrt{r_+}(2v\sqrt{r_+}-\sqrt{v^2+3}\sqrt{r_-})\big)^2,
 \\
 B&=&-\frac{l^2}{(r_+-r_-)^2}\big(2k+\o\sqrt{r_-}(2v\sqrt{r_-}-\sqrt{v^2+3}\sqrt{r_+})\big)^2,\\
 C&=&3(v^2-1)\o^2-m^2(v^2+3).
 \eea

The equation (\ref{radial}) could be transformed to standard
hypergeometric function form. Near the horizon $z=0$ or $r=r_+$,
there are two independent solutions
 \be
 \phi_1=z^\a (1-z)^\b F(a,b,c,z), \hspace{3ex} \phi_2=z^\a
 (1-z)^\b F(a+c-1, b-c+1, 2-c, z)
 \ee
 where
 \bea
 \a&=&-i\frac{\sqrt{A}}{v^2+3}, \nn\\
 \b&=&\frac{1}{2}\left(1-\sqrt{1-\frac{4Cl^2}{(v^2+3)^2}}\right), \nn
 \eea
 and
 \bea
 c&=&2\a+1,\nn\\
 a&=&\a+\b+i\sqrt{-B}/(v^2+3),\nn\\
 b&=&\a+\b-i\sqrt{-B}/(v^2+3).\nn
 \eea
 Since by definition the quasinormal modes have to be purely ingoing at the horizon,
 $\phi_1$ has the right boundary behavior. And in usual anti-de Sitter spacetime
 the boundary condition
 that at the infinity the  solution should vanish. For the warped $AdS_3$ spacetime,
 the boundary condition at infinity is still under investigation\cite{warped,AndyWei}.
 It is not natural to impose the vanishing Dirichlet boundary condition at radial
 infinity, since the conformal boundary of warped
 $AdS_3$ is very different from the one of usual $AdS_3$. Actually one
 may analyze the asymptotic differential equation for radial wavefunction
 and find that the potential is not infinitely high at radial
 infinity. Therefore, instead of choosing some kind of boundary
 condition on wavefunction directly at infinity, we impose the
 physical requirement  that the wavefunction is just purely
 outgoing at infinity and its corresponding flux is finite. The
 flux is \be \mathcal
{F}=\sqrt{-g}g^{rr}\frac{1}{2i}(\Phi^*\partial_r\Phi-\Phi\partial_r\Phi^*)\\
\propto (r-r_+)(r-r_-)(\Phi^*\partial_r\Phi-\Phi\partial_r\Phi^*)
\ee
  If $\omega$ is a ordinary complex number with both real and imaginary parts,
  then $\beta$ is a complex number and not real.
  We require Re$\sqrt{1-\frac{4Cl^2}{(v^2+3)^2}}\geq 0$ to avoid any
  ambiguity. So the asymptotic flux at the infinity has a set of
  divergent terms and the leading  term is of order $(1-z)^{2\beta-1}$,
  where Re$2\beta-1<0$.
  If $\omega$ is a purely imaginary number, then $\beta$ is real and $\beta<0$ and
  the
  leading order divergent term is $(1-z)^{2\beta}$.
The leading term are all proportional to
  \be
   \left |\frac{\Gamma(c)\Gamma(c-a-b)}{\Gamma(c-a)\Gamma(c-b)}\right |^2
  \ee
  Thus, if we require the asymptotic flux at the infinity is not divergent,
  we have to set
  \be
  c-a=-n, \hspace{5ex} \mbox{or}\hspace{5ex} c-b=-n,
  \ee
  with $n$ being a non-negative integer. Note that these two relations could
  be obtained by simply imposing vanishing Dirichlet condition at infinity.
  Let us analyze them case by
  case.

\begin{enumerate}
\item[1)]Case 1: $c-a=-n$\\
In this case, we are led to the following
equation on $\o$:
 \bea\label{case1}
 -i\frac{l}{r_+-r_-}\frac{1}{v^2+3}(4k+\o\d)
 +\frac{1}{2}\left(1+\sqrt{1-\frac{4Cl^2}{(v^2+3)^2}}\right)=-n,
 \eea
where
 \be
 \d\equiv 2v(r_++r_-)-2\sqrt{(v^2+3)r_+r_-}.
 \ee
The solution of (\ref{case1}) is of quite involved form:
 \bea\label{oR}
 \o_R&=&\frac{v^2+3}{d^2\d^2-3(v^2-1)l^2}\left\{-d\d\left(\frac{4kd}{v^2+3}+i(n+\frac{1}{2})\right)
 -i(e+if)\right\},
 \eea
 where
 \be
 d\equiv \frac{l}{r_+-r_-}.
 \ee
 And in the above solution, we have introduced
 \be
 e=\sqrt{\frac{\sqrt{E^2+F^2}+E}{2}}, \hs{3ex}
 f=\sqrt{\frac{\sqrt{E^2+F^2}-E}{2}},
 \ee
where
 \bea
 E&=&(\frac{1}{4}+\frac{m^2l^2}{v^2+3})d^2\d^2-3(v^2-1)l^2
 \left((\frac{1}{4}+\frac{m^2l^2}{v^2+3})+(\frac{4kd}{v^2+3})^2-(n+\frac{1}{2})^2\right),\nn\\
 F&=&-3(v^2-1)l^2(n+\frac{1}{2})\frac{8kd}{v^2+3}. \nn
 \eea

\item[2)] Case 2: $c-b=-n$\\
In this case, the equation on $\o$ is much simpler,
 \be
 -n-\frac{1}{2}+i\frac{2vl\o}{v^2+3}=\frac{1}{2}\sqrt{1-\frac{4Cl^2}{(v^2+3)^2}},
 \ee
which has the solution
 \be
 \o_L=-i\frac{1}{l}\left\{(2n+1)v+\sqrt{3(n+\frac{1}{2})^2(v^2-1)+(\frac{1}{4}+\frac{m^2l^2}{v^2+3})(v^2+3)}\right\}.
 \ee
Note that the frequency is pure imaginary, being independent of
the angular momentum.

\end{enumerate}

The above expression looks quite involved. It would be
illuminating to consider the $v=1$ limit to get a favor and have a
consistent check. In the $v=1$ limit, we have
 \bea
 \o_R&=&-\frac{2k}{(\sqrt{r_+}-\sqrt{r_-})^2}-i
 \frac{2(\sqrt{r_+}+\sqrt{r_-})}{l(\sqrt{r_+}-\sqrt{r_-})}\left(n+\frac{1}{2}(1+\sqrt{1+m^2l^2})\right)\\
 \o_L&=&-i\frac{2}{l}\left(n+\frac{1}{2}(1+\sqrt{1+m^2l^2})\right).
 \eea
At first looking, the above result is slightly different from
(\ref{pole}) obtained in \cite{Birmingham01}. The differences
reside on the temperature and angular momentum. Both differences
originate from the choice of different coordinate systems. For
example, taking into account of the coordinate difference, the
temperature should be as (\ref{tempr}), so that we still have the
same physical picture $\o_{L,R} \propto -\frac{4\pi i}{l}
\tilde{T}_{L,R}(1+h_{L,R})$.

 The above solution could be simplified in the following limits.
\begin{enumerate}
\item In the limit of $v\to \infty$, the expression of the
solutions could be simplified to
 \bea
 \o_R&=&-i\frac{v}{d^2\tilde{\d}^2-3l^2}\left(d\tilde{\d}(n+\frac{1}{2})+
 \sqrt{\frac{1}{4}d^2\tilde{\d}^2+3l^2n(n+1)}\right),\\
 \o_L&=&-i\frac{v}{l}\left(n+\frac{1}{2}+\sqrt{3(n+\frac{1}{2})^2+1}\right)
 \eea
 where
 \be
 \tilde{\d}=2({r_+}+{r_-}-\sqrt{r_+r_-}).
 \ee
The real part of the quasinormal modes and the mass dependence are
suppressed in the large $v$ limit. The quasinormal frequencies
look pure imaginary and are proportional to $v$. Physically this
means that since the spacetime is stretched so much,  the
relaxation of the system to thermal equilibrium is extremely fast.

\item For the extremal black hole with $r_+=r_-$, the fact that
$d\to \infty$ leads to
 \be
 \o_R=-\frac{4k}{\d}.
 \ee
Therefore, there is no imaginary part no matter how large $v$ is.
This may indicates that the right temperature is zero. And the
left temperature could be taken as the constant $\frac{v}{2\pi}$.
We note that this phenomenon is similar to the case in Kerr/CFT
correspondence\cite{AndyWei}, where $T_R=0,T_L=1/2\pi$.

Another subtle point arise in the extremal BTZ black hole. From
our calculation, it seems that $\tilde{T}_R$ in (\ref{tempr})
diverge. However the temperatures in (\ref{tempBTZ}) is finite.
This indicates that the coordinate transformation is singular and
the definition of the temperature is closely related to the choice
of the coordinates\cite{Kim08}.

\end{enumerate}

Let us return to the general case. The expression of the right
frequency looks awful. To simplify the discussion, we may let the
angular momentum vanishing $k=0$. Then we have
 \bea
 \o_R&=&-i\frac{v^2+3}{d^2\d^2-3(v^2-1)l^2}\left\{(n+\frac{1}{2})d\d \right.\nn\\
  & &\left. +\frac{1}{2}
 \sqrt{(1+\frac{4m^2l^2}{v^2+3})\left(d^2\d^2-3(v^2-1)l^2\right)+3(2n+1)^2(v^2-1)l^2}\right\}.
 \eea
With the expression of $\o_L$, it seems that the quasinormal
frequencies is not in agreement with the prediction (\ref{pole})
from AdS/CFT correspondence.

However, we have a few remarks:
\begin{itemize}
\item One interesting point is that in the above relations, there
is a combination
 \be
 1+\frac{4m^2l^2}{v^2+3}\nn
 \ee
 in the squared root. This reminds us of the conformal weight of
 the scalar field with mass $m$ in anti-de Sitter spacetime with
 radius $2l/\sqrt{v^2+3}$.

\item If we can neglect the $\o^2$ term in $C$, we  have
$C=-m^2(v^2+3)$ so that
 \bea\label{olow}
 \o_L&=&-i\frac{1}{l}\frac{v^2+3}{2v}\left(n+\frac{1}{2}\tilde{\Delta}\right),\\
 \o_R&=&-\frac{4k}{\d}-i\frac{v^2+3}{l}\frac{r_+-r_-}{\d}
   \left(n+\frac{1}{2}\tilde{\Delta}\right),
 \eea
with
 \be
 \tilde{\Delta}= 1+\sqrt{1+\frac{4m^2l^2}{v^2+3}}.
 \ee
   This is actually in exact match with
the prediction (\ref{pole}) of $AdS_3/CFT_2$ correspondence.
Firstly note that $\tilde{\Delta}$ is the scaling dimension of the
dual operator corresponding to the scalar field of mass $m$ in
$AdS_3$ spacetime with radius $2l/\sqrt{v^2+3}$. Secondly note
that for scalar perturbation in
 $CFT$, one has $h_L=h_R=\tilde{\Delta}/2$. Thirdly the temperature has to be
modified with respect to coordinate transformation. In
\cite{Andy08}, the temperatures are identified as
 \be
 T_L=\frac{1}{2\pi l}\frac{v^2+3}{8v}\d, \hspace{5ex}
 T_R=\frac{1}{2\pi l}\frac{v^2+3}{4}(r_+-r_-).
 \ee
And the temperatures in the quasinormal frequencies are
 \be
 \tilde{T}_{L,R}=T_{L,R}/\d.
 \ee
This is what also happens in BTZ case.

We need to justify our approximation. In order to have
$3(v^2-1)\o^2<<m^2(v^2+3)$, one may expect that in the very low
frequency limit $|\o|<<m$, the above approximation make sense.
However, the expression  (\ref{olow}) could not be in consistent
with this limit. The other possibility is when $v$ is very close
to $1$. In this case, the black hole solutions are very much like
BTZ black holes. In a sense, the BTZ black hole is the only one
which can exactly match the AdS/CFT prediction. A slightly
deviation from BTZ black hole is still in good agreement with
AdS/CFT correspondence.

\item There is a singular point at which the right-moving
quasinormal mode is not well-defined. Notice that
 \be
 d^2\delta^2-3(v^2-1)l^2=\frac{l^2}{(r_+-r_-)^2}(\sqrt{v^3+3}(r_++r_-)-4v\sqrt{r_+r_-})^2,
 \ee
 which vanishes when
 \be
 \frac{r_-}{r_+}=
 \frac{2v-\sqrt{3(v^2-1)}}{2v+\sqrt{3(v^2-1)}}.
 \ee
This relation is in consistent with the fact that $0<r_-/r_+\leq
1$. However, in this case, from (\ref{oR}) the right-moving
quasinormal mode has a negative infinite imaginary part,
indicating its damping time is zero. The AdS/CFT correspondence
suggest that the dual right temperature is infinite, which is not
true even after taking into account of the redefinition of the
temperature due to coordinate choice. Note that this issue does
not arise in the $v=1$ limit.

 \item The discrepancy between semi-classical gravity
calculation and the prediction of AdS/CFT correspondence may
originate from the deformation of $AdS_3$ spacetime at asymptotic
region. The warped $AdS_3$ black hole is asymptotic to the warped
$AdS_3$ spacetime, in which case the AdS/CFT correspondence has
not been well understood. For example, we have no idea of the
exact relation between the masses of scalar fields in the bulk and
the conformal weight $(h_L,h_R)$ of corresponding operators in
CFT. The usual way to decide the relation from supergravity
calculation cannot be applied here since the boundary of the
warped $AdS_3$ is not the spacetime in which the $CFT_2$ is
defined. Furthermore, our study suggest that if there does exist a
warped AdS/CFT correspondence, the identification of the
quasinormal modes of the warped $AdS_3$ blackhole with the poles
in the retarded Green funciton in the dual field theory may not be
exact. Obviously a further thorough study of warped AdS/CFT
correspondence  is deserving.
\end{itemize}

\section*{Acknowledgments}

We would like to thank Wei Song for very valuable discussions.
 The work was partially supported by NSFC Grant
No.10535060,10775002 and NKBRPC (No. 2006CB805905).
 \ed